\documentclass[12pt]{article}
\usepackage{amssymb,amsmath,epsfig}
\usepackage{graphicx}
\allowdisplaybreaks
\begin{document}

\title{\bf Complexity Factor for Static Cylindrical System in Energy-momentum Squared Gravity}

\author{M. Sharif \thanks{msharif.math@pu.edu.pk} and Ayesha
Anjum \thanks{ayeshaanjum283@gmail.com}\\
Department of Mathematics and Statistics, The University of Lahore,\\
1-KM Defence Road Lahore, Pakistan.}

\date{}

\maketitle
\begin{abstract}
This paper investigates some physical features that give rise to
complexity within the self-gravitating static cylindrical structure
coupled with anisotropic distribution in the energy-momentum squared
gravity. To accomplish this, we formulate the modified field
equations and explore the structure of the astronomical body. The
C-energy and Tolman mass are also calculated to discuss the matter
distribution. We then obtain some structure scalars via orthogonal
splitting of the Riemann tensor. Since, the complexity of the
considered structure is influenced by a variety of variables,
including anisotropic pressure and inhomogeneous energy density,
etc. thus, we adopt the factor $\mathcal{Y}_{TF}$ as the complexity
factor. Further, the complexity-free condition along with the
Gokhroo-Mehra model and polytropic equation of state are taken to
generate their corresponding solutions. We deduce that the inclusion
of additional terms of this modified theory leads to a more
complicated system.
\end{abstract}

\section{Introduction}

Cosmic structures developed entirely under the effect of gravity
result in the formation of self-gravitating systems (like stars,
planets, galaxies and stellar clusters). Numerous astronomical
observations \cite{1a} demonstrate that such large scale self-gravitating
systems provide crucial details about the beginning and evolution of
our universe. In order to comprehend the structure and formation of
the cosmos, it is significant to investigate these compact
structures. The physical characteristics of a complicated
astronomical structure may substantially change as a result of a
little disruption in the system. Therefore, it is necessary to
develop a complexity factor that links fundamental physical factors.
Furthermore, an adequate complexity factor must evaluate the impact
of both external and internal perturbations on the evolution and
stability of stellar configurations.

There is a large body of literature \cite{1} in developing a
suitable definition of complexity but a more appropriate
conventional definition has not been achieved. Entropy and
information have been considered in the earlier definitions of
complexity but could not accurately estimate complexity of two
standard models in physics: perfect crystal and ideal gas. Since
particles in perfect crystal are distributed in a specific manner
hence, minimum information is required to explain its symmetric
distribution. For the ideal gas, atoms are arbitrarily arranged,
therefore the maximum amount of data is required to determine any of
its potential states. These two systems exhibit contradictory
behavior but both give zero complexity. Since entropy and
information cannot adequately describe the complexity, so some other
parameters must be included in the definition of complexity.

The perception of complexity has been expanded by Lopez-Ruiz et al.
\cite{61} through analyzing the idea of disequilibrium, that
estimates the discrepancy between the equiprobable arrangement and
multiple probabilistic states of the system. In order to calculate
the complexity of compact objects like white dwarfs and neutron
stars, researchers have substituted the energy density of the system
in place of probability distribution \cite{62}. As particles are
compactly organized in the core of dense celestial bodies, thus
fewer radial pressure than tangential pressure is created, leading
to pressure anisotropy in the system. Consequently, anisotropy plays
a crucial role in evaluating the stability of self-gravitating
structures. The concept of complexity defined by Lopez-Ruiz and his
collaborators did not propose productive criteria to determine
complexity because it focused solely on energy density while
ignoring other state parameters like anisotropic pressure.

In the framework of general relativity (GR), Herrera \cite{5}
recently introduced a new approach to estimate the complexity factor
for a static sphere. In this technique, he assumed that the simplest
system has isotropic pressure and homogenous energy density, i.e.,
the complexity of such a system is zero. He connected these state
parameters in a single frame by splitting the Riemann tensor
orthogonally and established a complexity factor. Herrera et al.
\cite{7} extended this concept by introducing minimal complexity
condition for a dynamical fluid distribution evolving in a
homologous pattern. Sharif and Butt \cite{6} investigated complexity
of a static cylindrical configuration in GR and concluded that the
two parameters, i.e., inhomogeneous energy density and anisotropy in
pressure make the system more complicated. The same authors
\cite{15} studied the complexity of the charged cylindrical
structure and deduced that the electromagnetic field can create more
complexity in a self-gravitating structure. A static axially
symmetric distribution has also been examined by evaluating three
complexity factors \cite{8}. Herrera et al. \cite{9} studied a
quasi-homologous system with complexity-free criteria.

Cylindrical systems have been discussed at various scales to study
their behavior and role of different physical characteristics
\cite{A}. Levi-Civita \cite{9a} developed vacuum solutions (having
two independent components), which motivated researchers to
investigate relativistic phenomena and the underlying mysteries of
various celestial bodies. Einstein and Rosen \cite{10} obtained
solutions of the cylindrical gravitational waves. The presence of
naked singularity producing strong gravitational waves at the end of
collapse provided motivation to examine other physical properties of
cylindrical structures. Herrera and Santos \cite{11} determined the
matching conditions for dynamical collapsing cylindrical structure.
They demonstrated that the radial pressure exists at the surface of
the cylinder and its temporal component depends on the collapsing
matter. Sharif and Abbas \cite{12} explored the dynamics of
gravitational collapse of charged non-adiabatic cylinder and
investigated the influence of charge as well as heat on its
gravitational mass.

The evolution of the universe is effectively explained by GR through
the $\Lambda$ Cold Dark Matter model. However, there are certain
issues with this model namely the coincidence problem and
fine-tuning \cite{B}. In order to address the issues related to
cosmic expansion, many researchers introduced different extended
theories such as $f(R), f(R, T)$, where $R$ is the Ricci scalar and
$T$ denotes trace of the energy-momentum tensor $T_{\delta\lambda}$,
etc. The first generalization to GR was $f(R)$ theory, which is
established by substituting the generic function $f(R)$ in the
Einstein-Hilbert action in place of $R$ \cite{21}. Harko et al.
\cite{22} proposed an extension of $f(R)$ gravity, termed as
$f(R,T)$ through the use of gravitational Lagrangian density in the
form of $R$ and $T$. The cosmic accelerated expansion and the
interaction between dark matter/dark energy are effectively
described by the curvature-matter coupling scenarios in $f(R,T)$
gravity \cite{22aa}. Haghani et al. \cite{60} proposed $f(R,T,Q)$
theory, where $Q=R_{\delta\lambda}T^{\delta\lambda}$, by defining a
strong dependence of geometry and matter distribution. A thorough
analysis of several standard problems and the most recent progress
of modified theories in cosmology is provided in \cite{22a}. The
complexity condition for both static and dynamic anisotropic matter
configurations has been examined in $f(R)$ scenario \cite{23}. Abbas
and Ahmad \cite{25} explored the complexity of several compact stars
in $f(R,T)$ theory and deduced that the complexity is minimal near
the surface. Several people \cite{27} used Herrera's concept of
complexity in the framework of these extended theories.

Katirci and Kavuk \cite{31} recently presented a new theory which
generalized GR by describing a particular coupling between matter
and gravity through a factor $T^{\delta\lambda}T_{\delta\lambda}$.
This theory is known as the energy-momentum squared gravity (EMSG)
or $f(R,\textbf{T}^{2})$ gravity with
$\textbf{T}^{2}=T^{\delta\lambda}T_{\delta\lambda}$. The predictions
of GR regarding singularities (like the big bang singularity) at
higher energy levels are no longer applicable due to expected
quantum fluctuations. In this context, EMSG is regarded as a
valuable framework since it addresses the big bang singularity by
supporting regular bounce having the least scale factor and finite
maximum energy density in the beginning of the universe. The
conservation law does not hold in this theory due to the coupling
between matter and geometry, which implies the existence of some
additional force. Consequently, the motion of test particles diverges
from the standard geodesic trajectory. Numerous astrophysical and
cosmological phenomena have been investigated in this theory.

Roshan and Shojai \cite{34} determined an exact solution of the
modified field equations and verified the probability of bounce at
early time by discussing isotropic and homogeneous distribution.
Using several cosmological models of this theory, Broad and Barrow
\cite{35} found a variety of exact solutions for an isotropic cosmos
and analyzed their behavior for the accelerated expansion, early and
late-time evolution, and the presence or absence of singularities.
Many astrophysical systems, such as neutron stars have been explored
by using particular model like $f(R,\textbf{T}^{2})=R+\chi
T^{\delta\lambda}T_{\delta\lambda}$, $\chi$ is the model parameter
\cite{32}. In order to discuss the current cosmic expansion,
Bahamonde et al. \cite{36} investigated the dynamical
characteristics of two distinct $f(R,\textbf{T}^{2})$ models. Sharif
and Gul \cite{37} examined this theory through the Noether symmetry
approach and investigated some feasible cosmological models. They
also investigated the dynamics associated with the cylindrical
collapse in the presence of electromagnetic field and dissipative
matter, reaching the conclusion that charge, dissipative matter and
modified parameters decrease the collapse rate \cite{38}. Recently,
we have discussed the complexity of charged static sphere in
$f(R,\textbf{T}^{2})$ scenario and concluded that the
electromagnetic field reduces the complexity of a spherical system
\cite{38a}.

The purpose of this article is to establish the complexity factor
for a static cylindrical distribution within $f(R,\textbf{T}^{2})$
framework. The paper is organized in the following manner. The
modified field equations for anisotropic fluid configuration are
derived in the next section. Section \textbf{3} addresses certain
physical characteristics of matter distribution. The structure
scalars are then developed in section \textbf{4}. We construct
complexity-free constraint in section \textbf{5} to generate
solutions of the EMSG field equations corresponding to a particular
form of energy density provided by Gokhroo-Mehra and polytropic
equation of state. Lastly, we summarize the main outcomes in section
\textbf{6}.

\section{The $f(R,\textbf{T}^{2})$ Field Equations}

The general Einstein-Hilbert action of $f(R,\textbf{T}^{2})$ gravity
is given by the following expression \cite{31}.
\begin{equation}\label{1}
I_{f(R,\textbf{T}^{2})}=\int\mathcal{L}_m\sqrt{-g}~
d^{4}x+\int\frac{f(R,\textbf{T}^{2})}{2\kappa^{2}}\sqrt{-g}~d^{4}x,
\end{equation}
where $\kappa^{2},~\mathcal{L}_m$ and $g$ are the coupling constant,
matter Lagrangian and determinant of the metric tensor
$g_{\delta\lambda}$, respectively. The energy-momentum tensor is
related to the Lagrangian density as follows
\begin{equation*}
T_{\delta\lambda}=2(-\frac{\partial \mathcal{L}_m}{\partial
g^{\delta\lambda}}+\frac{1}{2}g_{\delta\lambda}\mathcal{L}_m).
\end{equation*}
Varying Eq.(\ref{1}) with respect to $g_{\delta\lambda}$, we obtain
the EMSG field equations as
\begin{equation}\label{2}
\Theta_{\delta\lambda}f_{\textbf{T}^{2}}+R_{\delta\lambda}f_R+g_{\delta\lambda}\Box
f_R-\nabla_{\delta}\nabla_{\lambda}f_R-
\frac{1}{2}g_{\delta\lambda}f=\kappa^{2}T_{\delta\lambda},
\end{equation}
here, $R_{\delta\lambda}$ represents the Ricci tensor. Also,
$f_R=\frac{\partial f}{\partial
R},~f_{\textbf{T}^{2}}=\frac{\partial f}{\partial
\textbf{T}^{2}},~\Box=\nabla^{\delta}\nabla_\delta$, while
$\Theta_{\delta\lambda}$ is given as
\begin{eqnarray}\nonumber
\Theta_{\delta\lambda}=\frac{\delta \textbf{T}^{2}}{\delta
g^{\delta\lambda}}=\frac{\delta(T_{\delta\lambda}T^{\delta\lambda})}{\delta
g^{\delta\lambda}}&=&-2(\mathcal{L}_m+\frac{1}{2}T)T_{\delta\lambda}+\mathcal{L}_m
g_{\delta\lambda}T\\\label{3}&-&4\frac{\partial^{2}\mathcal{L}_m}{\partial
g^{\delta\lambda}\partial g^{\beta\alpha}}T^{\beta\alpha}+2
T^{\beta}_\delta T_{\lambda\beta}.
\end{eqnarray}

The energy-momentum tensor describing anisotropic distribution of
matter inside the cylinder is expressed as
\begin{equation}\label{4}
T^{\delta}_\lambda=\Pi^{\delta}_\lambda+\rho\emph{v}^{\delta}\emph{v}_\lambda-\emph{p}\emph{h}^{\delta}_\lambda,
\end{equation}
where $ \emph{p},~\Pi^{\delta}_\lambda,~\rho$ and
$\emph{v}^{\delta}$ denote the pressure, anisotropic tensor, energy
density and four-velocity, respectively. These terminologies are
described by the following expressions
\begin{eqnarray*}
\Pi^{\delta}_\lambda=\frac{\Pi}{3}(\emph{h}^{\delta}_\lambda+3\emph{s}^{\delta}\emph{s}_\lambda),\quad
\Pi=-\emph{p}_\bot+\emph{p}_r,\quad\emph{h}^{\delta}_\lambda=-\emph{v}^{\delta}\emph{v}_\lambda+\delta^{\delta}_\lambda,~
\emph{p}=\frac{1}{3}(2\emph{p}_\bot+\emph{p}_r ),
\end{eqnarray*}
where the tangential and radial pressures are $\emph{p}_\bot$ and
$\emph{p}_r$, respectively. The four-vector and four-velocity are
given as
\begin{eqnarray*}
\emph{s}^{\delta}=(0,\frac{1}{\mathcal{G}},0,0),\quad\emph{v}^{\delta}=(\frac{1}{\mathcal{F}},0,0,0),
\end{eqnarray*}
satisfying $\emph{v}_\delta\emph{v}^{\delta}=1,~
\emph{s}_\delta\emph{s}^{\delta}=-1,~\emph{v}_\delta\emph{s}^{\delta}=0.$
Different matter Lagrangians generate different field equations
because such Lagrangian has no precise definition. It is noticed
that $\mathcal{L}_m=\rho$ and $-\emph{p}$ are the most extensively
employed choices in the literature. In GR, these options are not
problematic but in the non-minimal coupling, these choices lead to
different outcomes \cite{39}. Thus, for the sake of convenience, we
take $\kappa=1$ and $\mathcal{L}_m=\rho$ which yields \cite{40}
\begin{eqnarray}\nonumber
\Theta_{\delta\lambda}&=&-2\rho(T_{\delta\lambda}-\frac{1}{2}g_{\delta\lambda}T)
-TT_{\delta\lambda}+2 T^{\beta}_\delta T_{\lambda\beta},
\\\label{5} \mathbb{G}_{\delta\lambda}&=&R_{\delta\lambda}-\frac{1}{2}R
g_{\delta\lambda}=\frac{1}{\kappa^{2}f_R}(T_{\delta\lambda}+T^{(C)}_{\delta\lambda})=T^{(D)}_{\delta\lambda},
\end{eqnarray}
where $\mathbb{G}_{\delta\lambda}$ is the Einstein tensor and the
modified terms (or the correction terms) of $f(R,\textbf{T}^{2})$
theory are denoted by $T^{(C)}_{\delta\lambda}$ and have the
following form
\begin{eqnarray}\nonumber
T^{(C)}_{\delta\lambda}&=&\frac{1}{f_R}\bigg\{\nabla_\delta
\nabla_\lambda f_R-g_{\delta\lambda}\Box f_R+g_{\delta\lambda}\left(
\frac{f-R\ f_R}{2}\right)\\\label{6} &-&\rho
g_{\delta\lambda}f_{\textbf{T}^{2}}T+
(T+2\rho)f_{\textbf{T}^{2}}T_{\delta\lambda}-2T^{\beta}_\delta
T_{\beta\lambda}f_{\textbf{T}^{2}}\bigg\}.
\end{eqnarray}

To investigate the compact structure, we assume a static
cylindrically symmetric spacetime confined by the hypersurface
($\Sigma$) as
\begin{equation}\label{7}
\emph{ds}^{2}=\mathcal{F}^{2}(r)\emph{dt}^{2}-\big(
\mathcal{G}^{2}(r)\emph{dr}^{2}+\mathcal{H}^{2}(r)\emph{d}\theta^{2}+\alpha^{2}\mathcal{H}^{2}(r)
\emph{d}z^{2}\big).
\end{equation}
Here, $\mathcal{F}, \mathcal{G}, \mathcal{H}$ are functions of $r$
and $\alpha$ is the arbitrary constant. The metric representing the external geometry is given as \cite{12}
\begin{equation}\label{7a}
\emph{ds}^{2}=-\frac{2\mathcal{M}}{\mathcal{R}}\emph{d}\nu^{2}+2\emph{d}\mathcal{R}\emph{d}\nu-\mathcal{R}^{2}(\emph{d}\theta^{2}+\alpha^{2}\emph{d}z^{2}),
\end{equation}where $\mathcal{M}$ and $\nu$ indicate the total mass in the exterior region and the retarded time, respectively. The necessary and sufficient constraints for the smooth matching of two metrics (\ref{7}) and (\ref{7a}) on the hypersurface are provided in \cite{12}. The spacetime (\ref{7}) can be made
analogous to general cylindrically symmetric distribution by
specifying $r$ in such a manner that coefficient of $\emph{d}z^{2}$
(i.e., $\mathcal{H}^{2}(r)$) equals to $r^{2}$. This conversion is
referred to as the tangential gauge and it transforms the metric (7)
into the following form
\begin{equation}\label{8}
\emph{ds}^{2}=\mathcal{F}^{2}(r)\emph{dt}^{2}-\big(
\mathcal{G}^{2}(r)\emph{dr}^{2}+r^{2}\emph{d}\theta^{2}+\alpha^{2}r^{2}
\emph{d}z^{2}\big).
\end{equation}
Taking covariant divergence of Eq.(\ref{2}), we have
\begin{equation}\label{9}
\nabla^{\delta}T_{\delta\lambda}=\frac{1}{\kappa^{2}}\nabla^{\delta}(\Theta_{\delta\lambda}f_{\textbf{T}^{2}})
-\frac{1}{2}g_{\delta\lambda}\nabla^{\delta}f,
\end{equation}
which indicates the non-conservation of energy-momentum tensor in
$f(R,\textbf{T}^{2})$ gravity implying the existence of an unknown
force which is responsible for the non-geodesic motion of particles
in celestial bodies. The modified field equations associated with
the spacetime (\ref{8}) are given as
\begin{eqnarray}\label{10}
-\frac{1}{r\mathcal{G}^{2}}\Bigg(\frac{1}{r}-\frac{2\mathcal{G}^{'}}{\mathcal{G}}\Bigg)&=&\frac{1}{f_R}\left(\rho+\varphi+
\varphi_{00}\right),\\\nonumber
\frac{1}{r^{2}\mathcal{G}^{2}}\Bigg(\frac{2r\mathcal{F}^{'}}{\mathcal{G}}+1\Bigg)
&=&\frac{1}{f_R}\left\{(\varphi_{11}-\varphi+\emph{p}_r)+\left(2\rho\emph{p}_r+\rho^{2}
\right.\right.\\\label{11}
&-&\left.\left.2\rho\emph{p}_\bot-2\emph{p}_r\emph{p}_\bot+\emph{p}^{2}_r\right)
f_{\textbf{T}^{2}}\right\},\\\nonumber
\frac{1}{r\mathcal{G}^{2}}\Bigg(\frac{r\mathcal{F}^{''}}{\mathcal{F}}-\frac{r\mathcal{F}^{'}\mathcal{G}^{'}}
{\mathcal{F}\mathcal{G}}-\frac{\mathcal{G}^{'}}{\mathcal{G}}+\frac{\mathcal{F}^{'}}{\mathcal{F}}\Bigg)&=&
\frac{1}{f_R}\left\{(\varphi_{22}-\varphi+\emph{p}_\bot
)+\left(\rho\emph{p}_\bot+\rho^{2}\right.\right.\\\label{12}
&-&\left.\left.\rho\emph{p}_r-\emph{p}_r\emph{p}_\bot\right)f_{\textbf{T}^{2}}\right\},
\end{eqnarray}
where
\begin{eqnarray}\nonumber
\varphi&=&\frac{1}{2}(f-Rf_R),\\\nonumber
\varphi_{00}&=&\frac{1}{\mathcal{G}^{2}}\left(\frac{\mathcal{G}^{'}}{\mathcal{G}}
-\frac{2}{r}\right)f^{'}_R-\frac{f^{''}_R}{\mathcal{G}^{2}},\\\nonumber
\varphi_{11}&=&-\frac{f^{'}_R}{\mathcal{G}^{2}}\bigg(\frac{\mathcal{F}^{'}}{\mathcal{F}}+\frac{2}{r}\bigg),\\\nonumber
\varphi_{22}&=&-\frac{f^{''}_R}{\mathcal{G}^{2}}-\frac{1}{\mathcal{G}^{2}}\left(\frac{\mathcal{F}^{'}}{\mathcal{F}}
-\frac{\mathcal{G}^{'}}{\mathcal{G}}+\frac{1}{r}\right)f^{'}_R,\\\nonumber
\end{eqnarray}
prime denotes derivative with respect to $r$.

\section{Physical Characteristics of Matter Distribution}

The C-energy formula \cite{41} is used to determine the matter
composition of the cylindrically symmetric structure. This is given
as
\begin{equation}\label{14}
\emph{m}(r)=\emph{l}~\mathbb{E}=\emph{l}\Big(\frac{1}{8}-\frac{1}
{8\emph{l}^{2}}\nabla_\delta\hat{r}\nabla^{\delta}\hat{r}\Big),
\end{equation}
where $\hat{r}=\mathcal{P}\emph{l}, \quad
\mathcal{P}^{2}=\psi_{(1)k}\psi^{k}_{(1)},\quad
\emph{l}^{2}=\psi_{(2)k}\psi^{k}_{(2)}.$ The quantities
$\mathcal{P}$ and $\emph{l}$ represent the circumference radius and
specific length, respectively. Also, $\mathbb{E}$ is the
gravitational energy per specific length,
$\psi_{(1)}=\frac{\partial}{\partial\theta}$ and
$\psi_{(2)}=\frac{\partial}{\partial z}$. The inner mass of the
considered distribution becomes
\begin{equation}\label{15}
\emph{m}(r)=\frac{r\alpha}{2}\big(\frac{1}{4}-\frac{1}{\mathcal{G}^{2}}\big)=
\frac{\alpha}{2}\bigg(\frac{r}{4}+\int^{r}_0\tilde{r}^{2}T^{0(D)}_0\emph{d}\tilde{r}\bigg).
\end{equation}
Utilizing Eqs.(\ref{10})-(\ref{12}) with (\ref{15}), the mass
function takes the form
\begin{eqnarray}\nonumber
\emph{m}&=&\Big(T^{0(D)}_0-T^{1(D)}_1+T^{2(D)}_2\Big)+\frac{\alpha
r}{8}-\frac{1}{\mathcal{G}^{2}}\bigg(\frac{\alpha
r}{2}+\frac{1}{r^{2}}\bigg)\\\label{16}
&-&\frac{1}{\mathcal{F}\mathcal{G}^{2}}\Big(\mathcal{F}^{''}-
\frac{\mathcal{F}^{'}\mathcal{G}^{'}}{\mathcal{G}}-\frac{\mathcal{F}^{'}}{r}+
\frac{\mathcal{F}\mathcal{G}^{'}}{r\mathcal{G}}+\frac{\mathcal{F}}{r^{2}}\Big).
\end{eqnarray}
Equation (\ref{11}) yields the value of
$\frac{\mathcal{F}^{'}}{\mathcal{F}}$ as
\begin{equation}\label{A}
\frac{\mathcal{F}^{'}}{\mathcal{F}}=\frac{\emph{m}}{\alpha
r^{2}}-\frac{r}{8}-\frac{r}{2}T^{1(D)}_1.
\end{equation}

The Riemann tensor determines the distortion of spacetime and is
expressed in terms of the Ricci scalar, the Weyl
($\mathcal{C}^\mu_{\delta\lambda\varepsilon}$) and Ricci tensors as
\begin{equation}\label{17}
R^{\mu}_{\delta\lambda\varepsilon}=\mathcal{C}^{\mu}_{\delta\lambda\varepsilon}+\frac{1}{2}R^{\mu}_\lambda
g_{\delta\varepsilon}-\frac{1}{2}R_{\delta\lambda}\delta^{\mu}_\varepsilon+
\frac{1}{2}R_{\delta\varepsilon}\delta^{\mu}_\lambda-\frac{1}{2}R^{\mu}_\varepsilon
g_{\delta\lambda}-\frac{1}{6} R\left(\delta^{\mu}_\lambda
g_{\delta\varepsilon}-\delta^{\mu}_\varepsilon
g_{\delta\lambda}\right).
\end{equation}
The tidal force on an object is determined by the Weyl tensor, which
is the trace-free part of the Riemann tensor. This can be separated
into electric ($\mathbb{E}_{\mu\nu}$) and magnetic
($\mathbb{H}_{\mu\nu}$) components using the observer's
four-velocity as
\begin{eqnarray}\nonumber
\mathbb{H}_{\delta\lambda}=\frac{1}{2}\eta_{\delta\alpha\varepsilon\beta}\mathcal{C}^{\varepsilon\beta}_{\lambda\sigma}
\emph{v}^{\alpha}\emph{v}^{\sigma},\quad
\mathbb{E}_{\delta\lambda}=\mathcal{C}_{\delta\gamma\lambda\sigma}\emph{v}^{\gamma}\emph{v}^{\sigma},
\end{eqnarray}
where
$g_{\alpha\iota\beta\sigma}=g_{\alpha\beta}g_{\iota\sigma}-g_{\alpha\sigma}g_{\iota\beta},\quad
\mathcal{C}_{\sigma\varepsilon\mu\nu}=(g_{\sigma\varepsilon\alpha\beta}g_{\mu\nu\pi\kappa}-
\eta_{\sigma\varepsilon\alpha\beta}\eta_{\mu\nu\pi\kappa})\emph{v}^{\alpha}\emph{v}^{\pi}\mathbb{E}^{\beta\kappa},$
and $\eta_{\delta\lambda\mu\nu}$ denotes the Levi-Civita tensor.
Different purely electric spacetimes including all static ones are
known in the literature [36]-[39]. Since the spacetime being
examined is of static nature, therefore the magnetic component
disappears. The electric part in terms of the projection tensor and
unit four-vector is written as
\begin{equation}\label{18}
\mathbb{E}_{\delta\lambda}=\frac{\mathcal{E}}{3}\left(3\emph{s}_\delta\emph{s}_\lambda+\emph{h}_{\delta\lambda}\right),
\end{equation}
where \begin{equation}\label{19}
\mathcal{E}=\frac{1}{2\mathcal{F}\mathcal{G}^{2}}\Big(\mathcal{F}^{''}-
\frac{\mathcal{F}^{'}\mathcal{G}^{'}}{\mathcal{G}}-\frac{\mathcal{F}^{'}}{r}+
\frac{\mathcal{F}\mathcal{G}^{'}}{r\mathcal{G}}+\frac{\mathcal{F}}{r^{2}}\Big),
\end{equation}
and its non-vanishing components are
$$\mathbb{E}_{11}=\frac{2}{3}\mathcal{E}\mathcal{G}^{2},\quad\mathbb{E}_{22}=
\frac{-1}{3}\mathcal{E}r^{2},\quad\mathbb{E}_{33}=\frac{-1}{3}\mathcal{E}\alpha^{2}r^{2},$$
with $\mathbb{E}^{\delta}_\delta=0$ and
$\mathbb{E}_{\mu\lambda}\emph{v}^{\lambda}=0$.

We would like to highlight that the electric component of the Weyl
tensor in Eq.(\ref{18}) is defined through a single scalar function
because of the constraints created by the Weyl gauge, however, it is
expressed in terms of two scalar functions for the general
cylindrical symmetric distribution. We establish a connection
between the mass function and the Weyl tensor to investigate several
properties of the cylindrical framework by using formulations of
C-energy and Tolman mass \cite{42}. Employing Eqs.(\ref{5}),
(\ref{15}) and (\ref{19}), we obtain the following relation
\begin{equation}\label{20}
\frac{\emph{m}}{\alpha
r^{3}}=\frac{1}{8r^{2}}-\frac{1}{6}\Big(T^{0(D)}_0-T^{1(D)}_1
+T^{2(D)}_2\Big)+\frac{\mathcal{E}}{3},
\end{equation}
yielding
\begin{equation}\label{21}
\mathcal{E}=\frac{1}{2r^{3}}\int^{r}_0\tilde{r}^{3}(T^{0(D)}_0)^{'}\emph{d}\tilde{r}-
\frac{1}{2}\Big(T^{1(D)}_1-T^{2(D)}_2\Big).
\end{equation}
Using Eq.(\ref{21}) in (\ref{20}), we obtain
\begin{eqnarray}\nonumber
\emph{m}(r)&=&\frac{1}{6}\int^{r}_0
\tilde{r}^{3}\bigg\{\left(\frac{1}{f_R}\right)^{'}(\rho+\varphi+\varphi_{00})+
\left(\frac{1}{f_R}\right)\big(\rho+\varphi
+\varphi_{00}\big)^{'}\bigg\}d\tilde{r}\\\label{22}&-&\frac{\alpha}{6}r^{3}T^{0(D)}_0+\frac{\alpha
r}{8}.
\end{eqnarray}
This demonstrates the connection between energy density
inhomogeneity and mass function in modified theory. The effects of
modified terms on the structural changes resulting from the
inhomogeneity of the energy density can be examined through the
above-mentioned equation. When the inward-directed force of
gravitation is counterbalanced by the outward pressure of a
celestial body, then the system is said to be in equilibrium. In GR,
the Tolman-Opphenheimer-Volkoff (TOV) equation is the counterpart of
the hydrostatic equilibrium equation. For anisotropic matter
distribution, we obtain the TOV equation in $f(R,\textbf{T}^{2})$
theory through Eq.(\ref{9}) as
\begin{eqnarray}\nonumber
\emph{p}^{'}_r&=&\frac{1}{(\emph{p}_r+2\rho-2\emph{p}_\bot)f_{\textbf{T}^{2}}+1}
\bigg[2\Big\{\big(\emph{p}^{'}_\bot(2\rho-7\emph{p}_\bot
+2\emph{p}_r)+\rho^{'}(2\emph{p}_\bot-\rho-2\emph{p}_r)\big)\\\nonumber
&+&\frac{\mathcal{F}^{'}}{\mathcal{F}}\Big(-\rho^{2}
+2\rho\emph{p}_\bot-
2\rho\emph{p}_r-\emph{p}^{2}_r-4\emph{p}^{2}_\bot
+2\emph{p}_r\emph{p}_\bot\Big)-\frac{2}{r}
\Big(3\rho\emph{p}_r-3\rho\emph{p}_\bot+\emph{p}^{2}_r\\\nonumber&-&
\emph{p}_r\emph{p}_\bot
+4\emph{p}^{2}_\bot\Big)\Big\}f_{\textbf{T}^{2}}
-\bigg\{\frac{\alpha(\rho+\emph{p}_r)}{r(\alpha
r-2\emph{m})}\Big[\frac{r^{3}} {f_R} \Big(
\big(2\rho\emph{p}_r+\rho^{2}-2\rho\emph{p}_\bot-
2\emph{p}_r\emph{p}_\bot\\\label{13}&+&\emph{p}^{2}_r\big)f_{\textbf{T}^{2}}-(\varphi-\emph{p}_r-\varphi_{11})\Big)
+\frac{\emph{m}}{\alpha}\Big]+\frac{2}{r}
(\emph{p}_r-\emph{p}_\bot)\bigg\}\bigg],
\end{eqnarray}

Tolman described another formula for the mass of a cylindrically
symmetric distribution having radius $r$ within the boundary
$\Sigma$ as \cite{42}
\begin{equation}\label{23}
\emph{m}_{Tol}=\frac{\alpha}{2}\int^{r}_0\tilde{r}^{2}\mathcal{F}\mathcal{G}\left(T^{0(D)}_0
-T^{1(D)}_1-2T^{2(D)}_2\right)\emph{d}\tilde{r}.
\end{equation}
Using the field equations, the above expression turns out to be
\begin{equation}\label{24}
\emph{m}_{Tol}=-\frac{\alpha\mathcal{F}^{'}r^{2}}{\mathcal{G}}.
\end{equation}
Inserting the value of $\mathcal{F}^{'}$ from Eq.(\ref{A}), the
Tolman mass is rewritten as
\begin{equation}\label{25}
\emph{m}_{Tol}=\frac{\mathcal{F}\mathcal{G}}{8}\Big(4\alpha
r^{3}T^{1(D)}_1-8\emph{m}+\alpha r\Big).
\end{equation}
In a static gravitational field, a test particle's gravitational
acceleration is associated with the Tolman mass as
\begin{equation*}
\emph{a}=\frac{\mathcal{G}^{-1}\mathcal{F}^{'}}{\mathcal{F}}=-\frac{\emph{m}_{Tol}}{\alpha\mathcal{F}r^{2}}.
\end{equation*}
This equation interprets the Tolman mass as the effective
gravitational mass. Equation (\ref{24}) can be written in a more
suitable way after some simplifications as \cite{43}
\begin{equation}\label{26}
\emph{m}_{Tol}=(\emph{m}_{Tol})_\Sigma\left(\frac{r}{\mathcal{R}}\right)^{3}-\alpha
r^{3}\int^{\mathcal{R}}_r
\frac{\mathcal{F}\mathcal{G}}{\tilde{r}}\bigg(\mathcal{E}-\frac{1}{2}(T^{1(D)}_1-T^{2(D)}_2)\bigg)\emph{d}\tilde{r},
\end{equation}
where $\mathcal{R}$ is the radius at the boundary. Using
Eq.(\ref{21}), the Tolman mass formula can be rewritten as
\begin{equation}\label{27}
\emph{m}_{Tol}=(\emph{m}_{Tol})_\Sigma\left(\frac{r}{\mathcal{R}}\right)^{3}+\alpha
r^{3}\int^{\mathcal{R}}_r
\frac{\mathcal{F}\mathcal{G}}{\tilde{r}}\bigg((T^{1(D)}_1-T^{2(D)}_2)
-\frac{1}{2r^{3}}\int^{r}_0
r^{3}(T^{0(D)}_0)^{'}\emph{d}r\bigg)\emph{d}\tilde{r}.
\end{equation}
This equation describes the Tolman mass with modified corrections
for static cylindrical symmetric spacetime that could be very
essential in identifying how the Weyl scalar $\mathcal{E}$,
inhomogeneity in the energy density and effective pressure
anisotropy interact.

\section{The Orthogonal Splitting of the Riemann Tensor}

Bel \cite{43s} was the first to investigate the orthogonal splitting
of the  Riemann tensor, establishing its left, right, and double
dual in accordance with standard fashion. All the information in the
Riemann tensor is contained in these tensors. Employing Bel's
approach, Herrera \cite{44} derived structures scalars which are a
collection of tensors representing in terms of some scalar
functions. There are various distinct features of such scalars.
First of all, they are scalars, which make complex systems easier to
deal with them than tensors. Additionally, this single tool
interacts with various system's components and provides a wide range
of information about the evolution of the structure including
expansion, inhomogeneity, shear evolution, etc. Using his method, we
take into account the following tensor quantities
\begin{eqnarray}\label{28}
\mathcal{Y}_{\delta\lambda}&=&R_{\delta\mu\lambda\nu}\emph{v}^{\mu}\emph{v}^{\nu},\\\label{29}
\mathcal{Z}_{\delta\lambda}&=&
^{*}R_{\delta\mu\lambda\nu}\emph{v}^{\mu}\emph{v}^{\nu}=
\frac{1}{2}\eta_{\delta\mu\gamma\varsigma}R^{\gamma\varsigma}_{\lambda\nu}\emph{v}^{\mu}\emph{v}^{\nu},\\\label{30}
\mathcal{X}_{\delta\lambda}&=&^{*}R^{*}_{\delta\mu\lambda\nu}\emph{v}^{\mu}\emph{v}^{\nu}
=\frac{1}{2}\eta^{\gamma\varsigma}_{\delta\mu}R^{*}_{\gamma\varsigma\lambda\nu}\emph{v}^{\mu}\emph{v}^{\nu}.
\end{eqnarray}
Here, $*$ represents the dual tensor which is defined as
$R^{*}_{\delta\lambda\kappa\omega}=\frac{1}{2}\eta_{\alpha\beta\kappa\omega}R^{\alpha\beta}_{\delta\lambda}$.
The Riemann tensor can be written by using the field equations in
(\ref{17}) as
\begin{equation}\label{31}
R^{\delta\alpha}_{\lambda\mu}=\mathcal{C}^{\delta\alpha}_{\lambda\mu}+2T^{(D)[\delta}_{[\lambda}\delta^{\alpha]}_{\mu]}
+T^{(D)}\left(\frac{1}{3}\delta^{\delta}_{[\lambda}\delta^{\alpha}_{\mu]}-
\delta^{[\delta}_{[\lambda}\delta^{\alpha]}_{\mu]}\right).
\end{equation}
Using the above expression, the Riemann tensor can be splitted as
\begin{equation}\nonumber
R^{\delta\alpha}_{\lambda\mu}=R^{\delta\alpha}_{\mathbb{(I)}\lambda\mu}+
R^{\delta\alpha}_{\mathbb{(II)}\lambda\mu}+
R^{\delta\alpha}_{\mathbb{(III)}\lambda\mu}.
\end{equation}
Here,
\begin{eqnarray}\nonumber
R^{\delta\alpha}_{\mathbb{(I)}\lambda\mu}&=&\frac{2}{f_R}\left[(\rho+\emph{p}_\bot)+\nabla^{[\delta}\nabla_{[\lambda}\delta^{\alpha]}_{\mu]}+
\left\{(\rho^{2}-\rho\emph{p}_r+\rho\emph{p}_\bot-\emph{p}_r\emph{p}_\bot)f_{\textbf{T}^{2}}\right\}\right.\\\nonumber
&\times&\left.\emph{v}^{[\delta}\emph{v}_{[\lambda}\delta^{\alpha]}_{\mu]}+
\left\{(\varphi-\emph{p}_\bot-\Box
f_R)+(-\rho^{2}+\rho\emph{p}_r-\rho\emph{p}_\bot+\emph{p}_r\emph{p}_\bot)f_{\textbf{T}^{2}}
\right\}\right.\\\nonumber
&\times&\left.\delta^{[\delta}_{[\lambda}\delta^{\alpha]}_{\mu]}+\left\{(\emph{p}_r-\emph{p}_\bot)+\left(\rho^{2}+3\rho\emph{p}_r-3\rho\emph{p}_\bot-
\emph{p}_r\emph{p}_\bot\right)f_{\textbf{T}^{2}}\right\}\emph{s}^{[\delta}\emph{s}_{[\lambda}\delta^{\alpha]}_{\mu]}\right],\\\nonumber
R^{\delta\alpha}_{\mathbb{(II)}\lambda\mu}&=&\frac{1}{
f_R}\left\{\emph{p}_r+2\emph{p}_\bot-\rho-4\varphi+3\Box
f_R-(\emph{p}^{2}_r-3\rho^{2}+4\emph{p}_r\emph{p}_\bot)f_{\textbf{T}^{2}}\right\}\\\nonumber
&\times&
\left(\frac{2}{3}\delta^{[\delta}_{[\lambda}\delta^{\alpha]}_{\mu]}\right),\\\nonumber
R^{\delta\alpha}_{\mathbb{(III)}\lambda\mu}&=&4\big(\emph{v}^{[\delta}\emph{v}_{[\lambda}\mathcal{E}^{\alpha]}_{\mu]}
-\frac{1}{4}\epsilon^{\delta\alpha}_\nu\epsilon_{\lambda\mu\pi}\mathcal{E}^{\nu\pi}\big),
\end{eqnarray}
where $\epsilon_{\delta\lambda\mu}\emph{v}^{\mu}=0$ and
$\eta_{\delta\mu\lambda\nu}=\emph{v}_\delta\epsilon_{\mu\lambda\nu}
$. The structure scalars are a combination of state variables that
are important in evaluating the complexity of stellar structure and
particularly useful in examining physical characteristics of the
system.

The Riemann tensor allows us to express
$\mathcal{Y}_{\delta\lambda}$, $\mathcal{X}_{\delta\lambda}$ and
$\mathcal{Z}_{\delta\lambda}$ in terms of matter variables. Further,
these tensors are the source of five structure scalars. It is
mentioned here that instead of five, there are eight structure
scalars that correspond to the general cylindrical symmetric case.
Since the scalar associated with $\mathcal{Z}_{\delta\lambda}$ does
not contain state variables that are necessary to calculate the
complexity, therefore, we only consider four scalars in this work.
The tensors $\mathcal{Y}_{\delta\lambda}$ and
$\mathcal{X}_{\delta\lambda}$ can be written in terms of their trace
$(\mathcal{X}_{T}=\mathcal{X}^{\delta}_\delta,~\mathcal{Y}_{T}=
\mathcal{Y}^{\delta}_\delta)$ and trace-free
$(\mathcal{X}_{TF},~\mathcal{Y}_{TF})$ parts as
\begin{eqnarray}\nonumber
\mathcal{X}_{\delta\lambda}&=&\frac{1}{3}\emph{h}_{\delta\lambda}\mathcal{X}_T+
\mathcal{X}_{TF}\left(\frac{\emph{h}_{\delta\lambda}}{3}+\emph{s}_\delta\emph{s}_\lambda\right),\\\nonumber
\mathcal{Y}_{\delta\lambda}&=&\frac{1}{3}\emph{h}_{\delta\lambda}\mathcal{Y}_T+
\mathcal{Y}_{TF}\left(\frac{\emph{h}_{\delta\lambda}}{3}+\emph{s}_\delta\emph{s}_\lambda\right).
\end{eqnarray}
In $f(R,\textbf{T}^{2})$ gravity, the trace-free and trace
components are given as
\begin{eqnarray}\label{32}
\mathcal{X}_{TF}&=&\frac{1}{2f_R}\Big[
\left(3\rho\emph{p}_r+\emph{p}^{2}_r-\emph{p}_r\emph{p}_\bot-3\rho\emph{p}_\bot\right)f_{\textbf{T}^{2}}
+\left(\emph{p}_r-\emph{p}_\bot\right)\Big]-\mathcal{E},\\\label{33}
\mathcal{Y}_{TF}&=&\frac{1}{2f_R}\Big[\left(3\rho\emph{p}_r+\emph{p}^{2}_r-
\emph{p}_r\emph{p}_\bot-3\rho\emph{p}_\bot\right)f_{\textbf{T}^{2}}-\left(\emph{p}_\bot-\emph{p}_r\right)\Big]+\mathcal{E},\\\label{34}
\mathcal{X}_T&=&-\frac{\varphi}{f_R}+\frac{3\Box f_R}{2
f_R}-\frac{11}{2f_R}(\emph{p}_r\emph{p}_\bot+\frac{2}{11}\rho)f_{\textbf{T}^{2}},\\\nonumber
\mathcal{Y}_T&=&-\frac{\varphi}{f_R}+\frac{3}{2f_R}\big(\emph{p}+3\Box
f_R+\frac{1}{3}\rho\big)+\frac{1}{2f_R}\left(3\rho^{2}+\emph{p}^{2}_r-6\emph{p}_r\emph{p}_\bot\right)f_{\textbf{T}^{2}},\\\label{35}
\end{eqnarray}
Equation (\ref{32}) shows that the scalar $\mathcal{X}_{TF}$
determines energy density inhomogeneity in fluid distribution. The
overall energy content of the system is evaluated via
$\mathcal{X}_T$ in the presence of correction terms, while
$\mathcal{Y}_T$ examines the influence of anisotropic stresses
caused by inhomogeneous density. Equations (\ref{27}) and (\ref{33})
can be used to interpret physical significance of the scalar
$\mathcal{Y}_{TF}$ as
\begin{equation}\label{36}
\emph{m}_{Tol}=(\emph{m}_{Tol})_\Sigma\left(\frac{r}{\mathcal{R}}\right)^{3}-\alpha
r^{3}\int^{\mathcal{R}}_r
\frac{\mathcal{F}\mathcal{G}}{\tilde{r}}\bigg(\mathcal{Y}_{TF}-\frac{1}{2f_R}(\varphi_{22}-\varphi_{11})\bigg)\emph{d}\tilde{r}.
\end{equation}
Equations (\ref{33}) and (\ref{36}) demonstrate that
$\mathcal{Y}_{TF}$ determines how inhomogeneous energy density,
non-linear $f(R,\textbf{T}^{2})$ terms and anisotropic pressure
affect the Tolman mass. The local anisotropic pressure in the
presence of modified corrections can be obtained by utilizing
Eqs.(\ref{32}) and (\ref{33}) as
\begin{eqnarray}\nonumber
\mathcal{X}_{TF}+\mathcal{Y}_{TF}&=&\frac{1}{f_R}\left[(3\rho\emph{p}_r+\emph{p}^{2}_r
-\emph{p}_r\emph{p}_\bot-3\rho\emph{p}_\bot)f_{\textbf{T}^{2}}-(\emph{p}_\bot-\emph{p}_r)
\right].
\end{eqnarray}

\section{The Complexity Factor}

Complexity in a celestial structure is developed by a variety of
factors. The electromagnetic field, inhomogeneity, heat dissipation,
viscosity and pressure anisotropy, etc. are the examples of such
factors. In general, the only framework with zero complexity is the
one that has isotropic pressure and homogenous energy density.
Anisotropic pressure, energy density inhomogeneity and dark source
terms of EMSG are responsible for creating complexity in the system
under consideration. The scalar $\mathcal{Y}_{TF}$ relates these
factors and also evaluates their impacts on the Tolman mass.
Therefore, $\mathcal{Y}_{TF}$ is an appropriate choice for the
complexity factor of the current setup. Here, $\mathcal{Y}_{TF}$ in
terms of state parameters is produced by substituting Eq.(\ref{21})
in (\ref{33}) as
\begin{eqnarray}\nonumber
\mathcal{Y}_{TF}&=&\frac{1}{2r^{3}}\int^{r}_0\tilde{r}^{3}(T^{0(D)}_0)^{'}\emph{d}\tilde{r}-
\frac{1}{f_R}\bigg\{(\emph{p}_\bot-\emph{p}_r)+\frac{1}{2}\big(\varphi_{11}-\varphi_{22}\big)\\\label{37}
&-&\left(\emph{p}^{2}_r+3\rho\emph{p}_r-3\rho\emph{p}_\bot\
-\emph{p}_r\emph{p}_\bot\right)f_{\textbf{T}^{2}}\bigg\}.
\end{eqnarray}

The set of field equations in EMSG comprises of five unknown
parameters
$\left(\emph{p}_r,~\rho,~\emph{p}_\bot,~\mathcal{F},~\mathcal{G}\right)$,
so we need additional conditions to get a solution. For this
purpose, the vanishing complexity factor is used to establish one
constraint which is obtained from Eq.(\ref{37}) as
\begin{eqnarray}\nonumber
\Pi&=&\frac{1}{(\emph{p}_r+3\rho)f_{\textbf{T}^{2}}+1}\bigg[\frac{\left(\varphi_{11}-\varphi_{22}\right)}{2}-\frac{f_R}
{2r^{3}}\int^{r}_0 \tilde{r}^{3}\\\label{38}
&\times&\Big\{\Big(\frac{1}{f_R}\Big)\left(\varphi+\rho+\varphi_{00}
\right)^{'}+\left(\frac{1}{f_R}\right)^{'}\left(\varphi+\rho+\varphi_{00}\right)\Big\}\emph{d}\tilde{r}\bigg].
\end{eqnarray}
For homogenous and isotropic matter distribution in GR, the
complexity factor disappears. On the other hand, in
$f(R,\textbf{T}^2)$ gravity, the complexity vanishes for isotropic
and homogeneous distribution if the system satisfies the following
condition
\begin{eqnarray}\nonumber
\frac{r^{3}}{f_R}\left(\varphi_{11}-\varphi_{22}\right)-\int^{r}_0
\tilde{r}^{3}\Big\{\Big(\frac{1}{f_R}\Big)\left(\varphi+\rho+\varphi_{00}
\right)^{'}
+\left(\frac{1}{f_R}\right)^{'}\left(\varphi+\rho+\varphi_{00}\right)\Big\}\emph{d}\tilde{r}=0.
\end{eqnarray}
Now, we examine the zero complexity constraint for a particular
$f(R,\textbf{T}^{2})$ model given as \cite{34}
\begin{equation}\label{39}
f\left(R,\textbf{T}^{2}\right)=R+\chi\textbf{T}^{2},
\end{equation}
which leads Eq.(\ref{38}) to
\begin{eqnarray}\nonumber
\Pi&=&\frac{1}{r^{3}(2+(6\rho+2\emph{p}_r)\chi)}\left[\int^{r}_0\tilde{r}^{3}\rho^{'}\emph{d}\tilde{r}
+\frac{\chi}{2}\int^{r}_0\tilde{r}^{3}\left(\emph{p}^{2}_r+
2\emph{p}^{2}_\bot+\rho^{2}\right)^{'}\emph{d}\tilde{r}\right].\\\label{40}
\end{eqnarray}
We still need a constraint to solve the field equations even after
applying the condition $\mathcal{Y}_{TF}=0$. To achieve this goal,
we employ the energy density of the Gokhroo-Mehra solution and the
polytropic equation of state to construct the relevant solutions.

\subsection{The Gokhroo-Mehra Solution}

To evaluate the solutions to the field equations corresponding to
anisotropic self-gravitating structure, Gokhroo and Mehra \cite{46}
took into the account a particular type of energy density. They
developed a model which describes the behavior of neutron star as
well as higher redshifts of several quasi-stellar configurations. We
use this specific form of energy density for the current
configuration to analyze how compact structures will behave when the
condition of vanishing complexity is applied \cite{46}. Thus, the
energy density is expressed as
\begin{equation}\label{41}
\rho=\rho_o\left(1-\frac{\mathcal{K}r^{2}}{\mathcal{R}^{2}}\right),
\end{equation}
where $\mathcal{K}\in (0,1)$ and $\rho_o$ is a constant. For the
assumed energy density, the mass function takes the form
\begin{eqnarray}\nonumber
\emph{m}(r)&=&\frac{\alpha}{2}\bigg[\frac{\rho_or^{3}}{3}\big(1-\frac{3\mathcal{K}r^{2}}{5\mathcal{R}^{2}}+\frac{\chi}{2}
+\frac{3\chi\mathcal{K}^{2}r^{4}}{14\mathcal{R}^{4}}-\frac{3\chi\mathcal{K}r^{2}}{5\mathcal{R}^{2}}\big)
+\frac{r}{4}\\\label{41x}&+&\frac{\chi}{2}\int^{r}_0
\tilde{r}^{2}(2\emph{p}^{2}_\bot+\emph{p}^{2}_r)\emph{d}\tilde{r}\bigg],
\end{eqnarray}
which gives the following form of the metric function $\mathcal{G}$
\begin{eqnarray}\nonumber
\frac{1}{\mathcal{G}^{2}}&=&\frac{\chi}{2}\bigg(\frac{6\varsigma\mathcal{K}r^{4}}{5\mathcal{R}^{2}}-\varsigma
r^{2}-\frac{3\varsigma\mathcal{K}^{2}r^{6}}{7\mathcal{R}^{4}}-\frac{1}{r}\int^{r}_0
\tilde{r}^{2}(2\emph{p}^{2}_\bot+\emph{p}^{2}_r)\emph{d}\tilde{r}\bigg)\\\label{41y}&+&\frac{3\varsigma\mathcal{K}r^{2}}{5\mathcal{R}^{2}}-r^{2}\varsigma,
\end{eqnarray}
$\varsigma=\rho_o/3$. From Eqs.(\ref{11}) and (\ref{12}), it is
evident that
\begin{eqnarray}\label{42}
\frac{1}{r^{2}\mathcal{G}^{2}}\Big(1+\frac
{r\mathcal{F}^{'}}{\mathcal{F}}-\frac{r^{2}\mathcal{F}^{''}}
{\mathcal{F}}+\frac{r^{2}\mathcal{F}^{'}\mathcal{G}^{'}}{\mathcal{F}\mathcal{G}}
+\frac{r\mathcal{G}^{'}}{\mathcal{G}}\Big)=\Pi\left\{(\emph{p}_r+3\rho)\chi+1\right\}.
\end{eqnarray}
In order to find the unknowns, we introduce new variables as
\begin{eqnarray}\nonumber
\mathcal{G}^{-2}=g(r),\quad\mathcal{F}^{2}(r)=
\emph{e}^{2\int\left(f(r)-\frac{1}{r}\right)\emph{d}r},
\end{eqnarray}
hence, Eq.(\ref{42}) reduces to
\begin{equation}\nonumber
g^{'}+2g\left[f+\frac{f^{'}}{f}+
\frac{2}{r^{2}f}-\frac{3}{r}\right]=-\frac{2}{f}\big[\Pi\left\{1+\chi(\emph{p}_r+3\rho)\right\}\big].
\end{equation}
Integration of the above expression provides the radial metric
function as
\begin{equation}\nonumber
\mathcal{G}^{2}(r)=\frac{f^{2}(r)\emph{e}^{\int\left(\frac{4}{f(r){r}^{2}}
+2f(r)\right)\emph{d}r}}{r^{3}\left[-2\int\frac{f(r)\emph{e}^{\int\left(\frac{4}{f(r){r}^{2}}
+2f(r)\right)\emph{d}r}\big(1+\Pi(r)[1+\chi(\emph{p}_r+3\rho)]\big)}{r^{6}}\emph{d}r+\mathcal{C}\right]},
\end{equation}
where $\mathcal{C}$ represents an integration constant. Hence, the
line element can be expressed in the form of $\Pi$ and $f(r)$ as
follows
\begin{eqnarray}\nonumber
\emph{ds}^{2}&=&\frac{-f^{2}(r)\emph{e}^{\int\left(\frac{4}{f(r){r}^{2}}
+2f(r)\right)\emph{d}r}\emph{dr}^{2}}{r^{3}\left[-2\int\frac{f(r)\emph{e}^{\int\left(\frac{4}{f(r){r}^{2}}
+2f(r)\right)\emph{d}r}\big(1+\Pi(r)[1+\chi(\emph{p}_r+3\rho)]\big)}{r^{6}}\emph{d}r+\mathcal{C}\right]}
\\\label{41}
&-&r^{2}\emph{d}\theta^{2}-r^{2}\alpha^{2}\emph{d}z^{2}+\emph{e}^{2\int\left(f(r)-\frac{1}{r}\right)\emph{d}r}\emph{dt}^{2}.
\end{eqnarray}

\subsection {The Polytropic Model with Complexity-free Condition}

In analyzing the internal structure of self-gravitating systems,
several physical parameters play significant role. However, some
factors are more important than others in examining the structure.
For the current scenario, it is helpful to use an equation of state
which accurately describes the relationship of key factors.
Anisotropic celestial structures have extensively been studied
through the polytropic equation of state, which expresses how energy
density and radial pressure are related to each other \cite{47a}.
The polytropic equation of state has the following form
\begin{equation}\label{57a}
\emph{p}_r=\mathbb{K}\rho^{\gamma}=\mathbb{K}\rho^{\frac{1+n}{n}},
\end{equation}
where $\gamma$, $n$ and $\mathbb{K}$ indicate the polytropic
exponent, polytropic index and polytropic constant, respectively. To
derive the dimensionless mass function and TOV equation, we apply
the following variables
\begin{eqnarray}\nonumber
\rho_o=\frac{\emph{p}_{ro}}{\tau},\quad&&\xi=r\mathcal{J},\quad
\mathcal{J}^{2}=\frac{\rho_o}{(2n+2)\tau},\quad\Upsilon^{n}(\xi)\rho_o={\rho},\\\label{57b}&&\Omega(\xi)
=\frac{2\mathcal{J}^{3}\emph{m}(r)}{\rho_o},
\end{eqnarray}
where the entities $\Upsilon,~\Omega,~\tau$ and $~\xi$ are
dimensionless. We obtain the dimensionless forms of Eq.(\ref{15})
and (\ref{13}) by inserting the above variables
\begin{align}\label{44}
&\frac{\mathit{d}\Omega}{\mathit{d}\xi}=\frac{\alpha\rho_o}{8(n+1)\tau}+\xi^{2}\Upsilon^{n}\Big\{1
+\frac{\chi}{2}
\Big(\rho_o\Upsilon^{n}+3\tau^{2}\rho_o\Upsilon^{n+2}+\frac{2\Pi^{2}}{\rho_o\Upsilon^{n}}-4\tau\Upsilon\Pi\Big)\Big\},\\\nonumber
&\tau\xi^{3}\Upsilon^{n+1}\Big\{1+\frac{\chi}{2\alpha}
\Big(\rho_o\Upsilon^{n}+3\Upsilon^{2}\rho_o\Upsilon^{n+2}+\frac{2\Pi^{2}}{\rho_o\Upsilon^{n}}-4\tau\Upsilon\Pi
\Big)\Big\}-\frac{\xi^{3}\chi}{\alpha\rho_o}\Big[\rho^{2}_o\Upsilon^{n}\\\nonumber&\times\big(\frac{1}{2}\Upsilon^{n}
+\frac{1}{2}\tau^{2}\Upsilon-\tau^{2}\Upsilon^{n+2}-\tau^{2}\Upsilon^{n+1}\big)+4\tau
\rho_o\Pi\Upsilon^{n+1}\Big]+\frac{\big(\alpha-\frac{2(n+1)\tau\Omega}{\xi}\big)}{(1+\tau\Upsilon)}\\\nonumber&\times
\bigg[\Big\{(-\rho_o\Upsilon^{n}-2\Pi-7\tau^{2}\Upsilon^{n+2}-\frac{8\Pi^{2}}
{\rho_o\Upsilon^{n}})\Big\}^{-1}
\Big\{\big(\xi^{2}\frac{\mathit{d}\Upsilon}{\mathit{d}\xi}\big)
\big(1+(2\Pi-\tau\rho_o\Upsilon^{n+1}\\\nonumber&+2\rho_o\Upsilon^{n})
2\chi\big)-\frac{2\chi\xi^{2}}{\tau(n+1)}\big[-\rho_on\Upsilon^{n-1}\frac
{\mathit{d}\Upsilon}{\mathit{d}\xi}
+2\tau\rho_o(n+1)\Upsilon^{n}\frac{\mathit{d}\Upsilon}
{\mathit{d}\xi}-2\frac{\mathit{d}\Pi}{\mathit{d}\xi}
\\\nonumber&-2n\Upsilon^{-1}\Pi\frac{\mathit{d}\Upsilon}{\mathit{d}\xi}
-12\tau^{2}\rho_o(n+1)\Upsilon^{n+1}\frac{\mathit{d}\Upsilon}{\mathit{d}\xi}+12\tau\Upsilon\frac{\mathit{d}\Pi}{\mathit{d}\xi}
+14\tau(n+1)\Pi\frac{\mathit{d}\Upsilon}
{\mathit{d}\xi}\\\nonumber&-\frac{14\Pi}{\rho_o\Upsilon^{n}}\frac{\mathit{d}\Pi}{\mathit{d}\xi}
+\frac{2}{\xi}\big(3\Pi-15\tau\Upsilon\Pi-\frac{8\Pi^{2}}{\rho_o\Upsilon^{n}}
-8\tau^{2}\rho_o\Upsilon^{n+2}\big)
\big]+\frac{2\Pi\xi\Upsilon^{-n}}{\tau\rho_o(n+1)}\Big\}\bigg]
\\\label{43}&+\frac{\Omega}{\alpha}.
\end{align}
Here we have three unknowns $\Pi,~\Upsilon,~\Omega$ and we use
complexity-free condition to evaluate a unique solution. In
dimensionless form, the zero complexity condition (\ref{40}) is
expressed as
\begin{eqnarray}\nonumber
&&\frac{2\xi}{\rho_on}
\big[1+(\tau\Upsilon+3)\rho_o\chi\Upsilon^{n}-\tau\chi\rho_o\Upsilon^{n+1}
+\chi\Pi\big]\frac{\mathit{d}\Pi}{\mathit{d}\xi}=\bigg[\frac{-2\xi\Pi}{\rho_on}
\big(3\rho_on\Upsilon^{n-1}\\\nonumber&&+\tau\rho_o(n+1)\Upsilon^{n}\big)\chi+\xi\Upsilon^{n-1}+\frac{\chi\xi}{n}
\big(3\tau^{2}\rho_o(1+n)\Upsilon^{2n+1}-2\tau(n+1)\Upsilon^{n}\Pi\\\label{45}
&&+\rho_on\Upsilon^{2n-1}\big)\bigg]
\frac{\mathit{d}\Upsilon}{\mathit{d}\xi}-\frac{6\Pi}{\rho_on}\left(1+(\tau\Upsilon+3)\rho_o\Upsilon^{n}\chi\right).
\end{eqnarray}
\begin{figure}\center
\epsfig{file=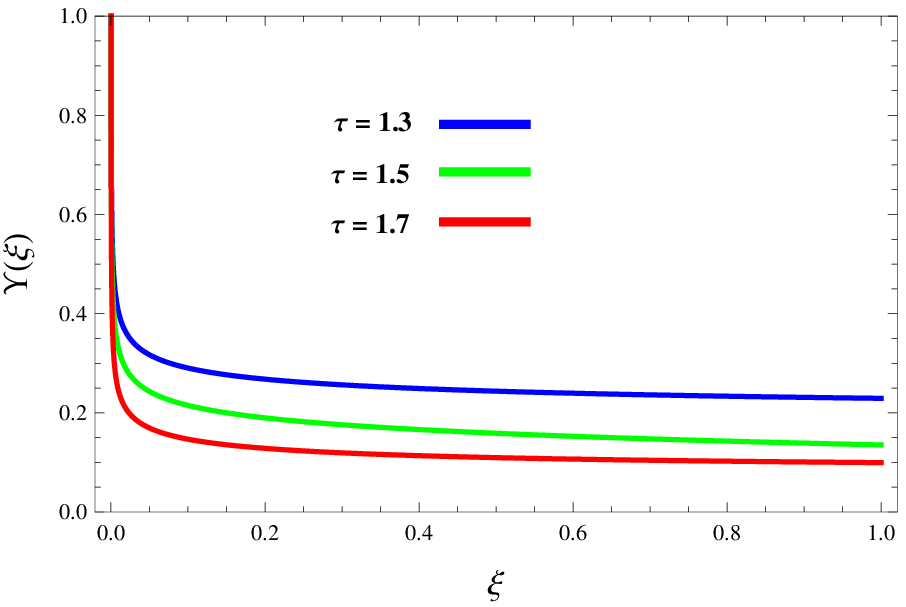,width=0.5\linewidth}\epsfig{file=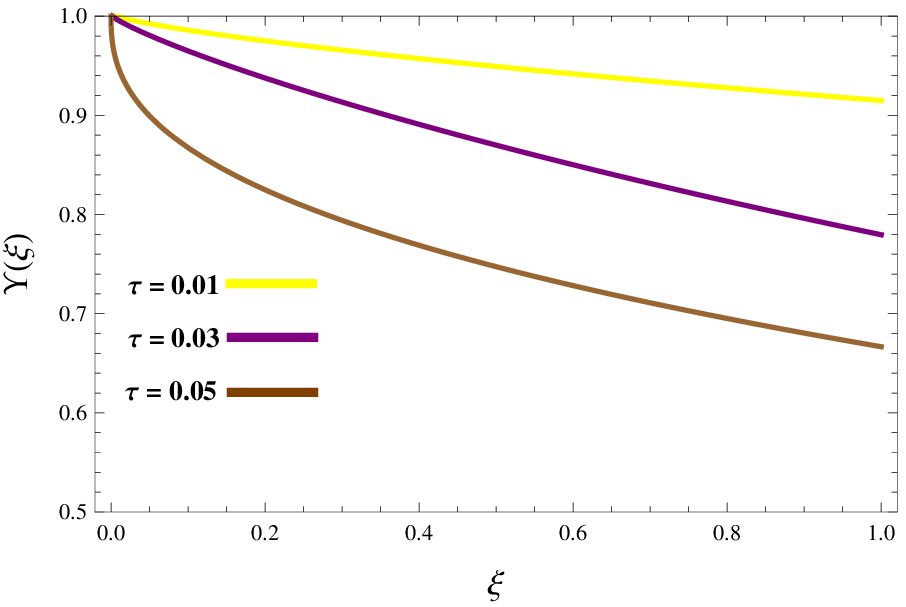,width=0.5\linewidth}
\caption{Plots of $\Upsilon$ versus $\xi$ for
$\chi=3,~n=5,~\alpha=1,~\rho_o=7$.}
\end{figure}
\begin{figure}
\epsfig{file=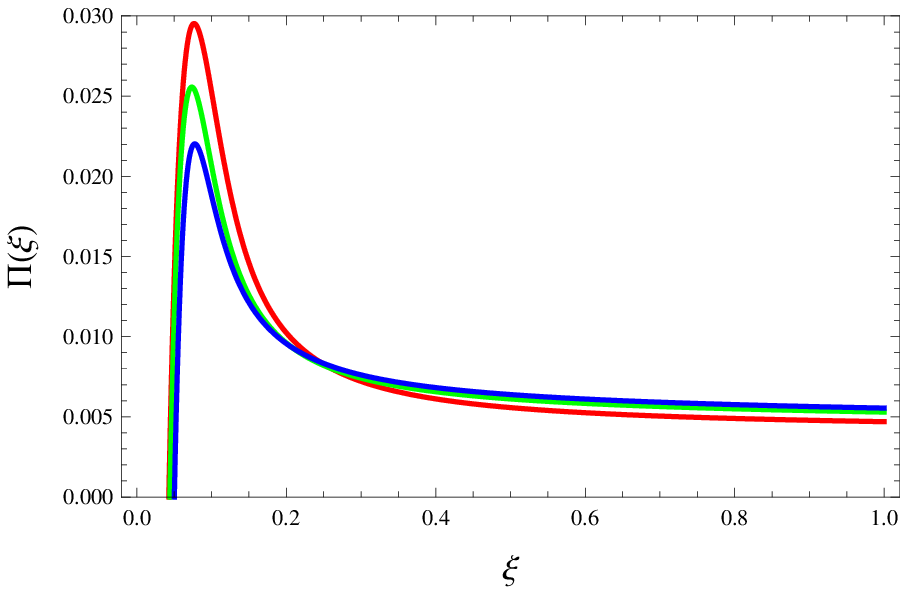,width=0.5\linewidth}\epsfig{file=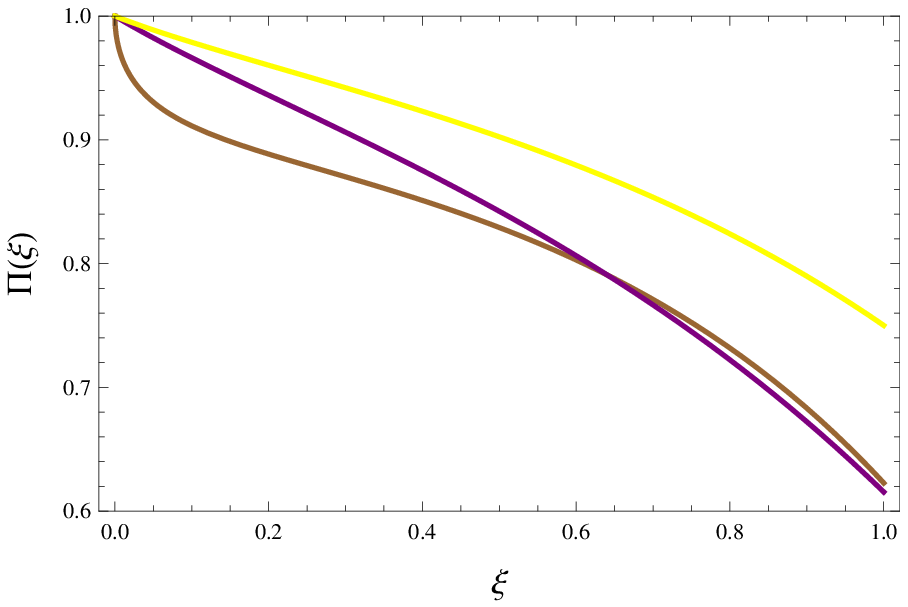,width=0.48\linewidth}
\caption{Plots of $\Pi$ versus $\xi$ for
$\chi=3,~n=5,~\alpha=1,~\rho_o=7$.}
\end{figure}

For some arbitrary values of $\tau$ and $n$, we obtain a unique
solution for the cylindrical celestial structure with zero
complexity. A physically acceptable model must have positive, finite
and maximum state parameters(pressure and energy density) at its
center $(r=0)$. Also, they must have decreasing trend towards the
boundary. Moreover, the mass function must be an increasing and
positive function of the radial coordinate. We fix $\chi=3, n=5,
\alpha=1$ and $\rho_o=7 $ for graphical analysis.  The behavior of
the energy density, anisotropy and mass function are illustrated
through Figures \textbf{1-3}, respectively. Figure \textbf{1} shows
that $\Upsilon$ is maximum at the center and exhibits decreasing
trend for smaller values of $\tau$. However, it shows a divergence
of the magnitudes as we approach to zero of the abscissa for larger
values of $\tau$ (left plot). Moreover, anisotropy also shows
diverging behavior near the center and then exhibits a rapid
increase followed by a decreasing trend towards the boundary for
$\tau=1.3, 1.5, 1.7$ (Figure \textbf{2}). However, this factor
exhibits decreasing behavior throughout for $\tau=0.01, 0.03, 0.05$
(right plot). Figure \textbf{3} indicates that the mass function
varies directly with $\xi$ while it has an inverse relation with
$\tau$. Since the left plots in Figures \textbf{1} and \textbf{2}
show a divergence, thus our resulting model is no more valid for
larger $\tau$. We conclude that this model has physically valid
solution only for $\tau=0.01, 0.03$ and $0.05$.
\begin{figure}
\epsfig{file=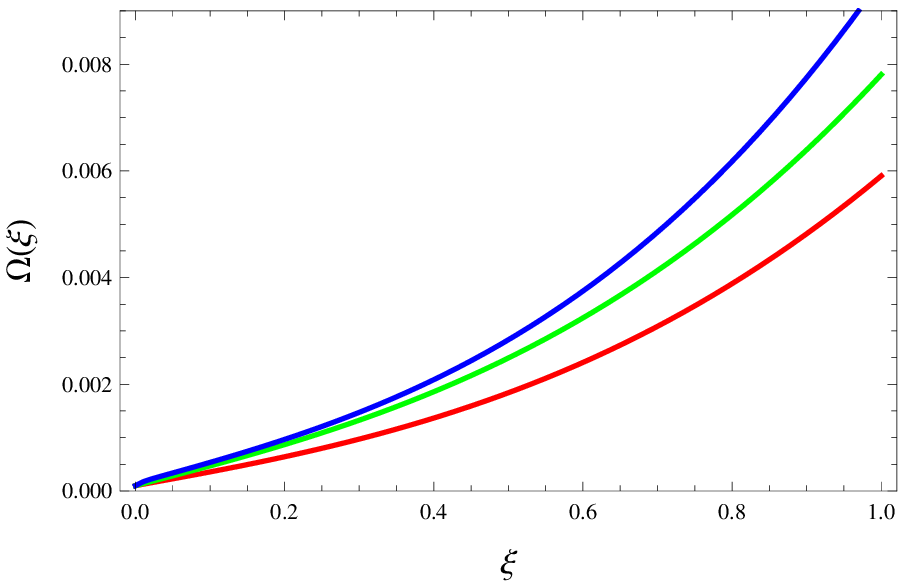,width=0.5\linewidth}\epsfig{file=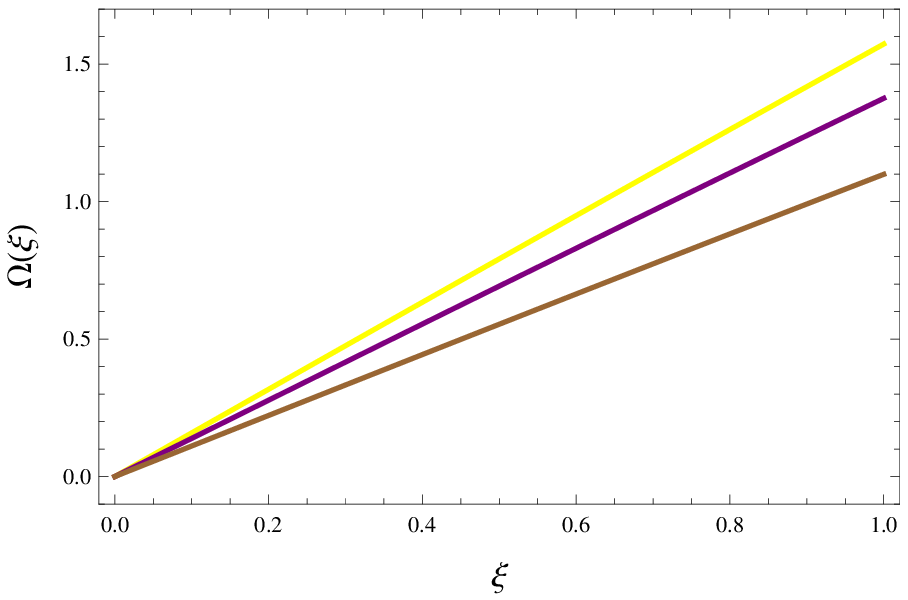,width=0.5\linewidth}
\caption{Plots of $\Omega$ versus $\xi$ for
$\chi=3,~n=5,~\alpha=1,~\rho_o=7$.}
\end{figure}

\section{Conclusions}

The presence of complexity in a structure identifies the existence
of pressure anisotropy and inhomogeneity in the energy density. A
physically uniform system in all directions has no complexity. In
this paper, we have analyzed the complexity of static cylindrical
object within the EMSG scenario. In this regard, the modified field
equations corresponding to anisotropic cylindrical distribution have
been computed. The mass functions $\emph{m}$ and $\emph{m}_{Tol}$
are calculated by using the C-energy and Tolman
formulations, respectively and a specific relationship between them
has also been developed. We have then discussed how the Weyl tensor and matter
variables are related to $\emph{m}_{Tol}$ and $\emph{m}$. There are
several definitions of complexity based on various factors in the literature.
Since the definition suggested by Herrera included all the factors which cause
complexity and thus, considers as the most suitable definition. Using
his method for the orthogonal splitting of the Riemann tensor, four structure scalars are obtained to formulate the complexity
factor. The scalar $\mathcal{Y}_{TF}$ incorporates the impacts of
all matter variables, including anisotropic pressure and
inhomogeneous energy density along with additional terms of
$f(R,\textbf{T}^{2})$ gravity. This scalar also deals with the
effects of inhomogeneous and anisotropic factors upon the Tolman
mass and, therefore, taken as the complexity factor.

The vanishing complexity condition has been constructed by assigning
$\mathcal{Y}_{TF}=0$. If a self-gravitating system in GR has an
isotropic and homogeneous configuration, then it is regarded as the
complexity-free. However, this does not imply vanishing complexity
in this theory, which reflects the influence of modified terms. Thus,
we can conclude that EMSG corrections are accounted for enhancing
the complexity of a cylindrical body. The complexity factor will be
zero if
\begin{eqnarray}\nonumber
\frac{r^{3}}{f_R}\left(\varphi_{11}-\varphi_{22}\right)-\int^{r}_0
\tilde{r}^{3}\Big\{\Big(\frac{1}{f_R}\Big)\left(\varphi+\rho+\varphi_{00}
\right)^{'}
+\left(\frac{1}{f_R}\right)^{'}\left(\varphi+\rho+\varphi_{00}\right)\Big\}\emph{d}\tilde{r}=0.
\end{eqnarray}
The complexity-free constraint for a particular model
$f\left(R,\textbf{T}^{2}\right)=R+\chi\textbf{T}^{2}$ gives an
additional constraint which helps in solving the field equations by
lessening the degrees of freedom.

Finally, two distinct models have been discussed to compute the
solutions of modified field equations. Utilizing the energy density
of the stellar configuration proposed by Gokhroo-Mehra, we have
explored characteristics of compact objects and obtained the
corresponding solution. For the second model, the polytropic
equation of state has been employed to establish a set of
dimensionless equations by incorporating certain new parameters. The
current setup consists of dimensionless zero complexity condition,
mass and TOV equation.
have calculated numerical solutions of this system and examined them
graphically by varying the parameter $\tau$. Figure \textbf{1}
demonstrates that the smaller values of $\tau$ provide maximum
density at the center and decreasing towards boundary while larger
values of $\tau$ yield divergence in its magnitude near the center.
Furthermore, for $\tau=1.3, 1.5$ and $1.7$, anisotropy also displays
a diverging trend near the core. However, this parameter
consistently shows decreasing behavior for $\tau=0.01, 0.03$ and
$0.05$. Figure \textbf{3} shows an inverse relationship between the
mass function and $\tau$, while it varies directly with $\xi$. The
graphs of $\Upsilon$ and $\Pi$ show a diverging behavior for larger
$\tau$, resulting in an invalid model. 
that our results for smaller values of $\tau$ provide physically
viable solution contrary to the charged \cite{38a} and uncharged
\cite{48} sphere in this theory. It can be observed from the
graphical analysis that $f(R,\textbf{T}^2)$ theory produces more
dense structure as compared to GR. All our results coincide with GR
\cite{6} for $f(R,\textbf{T}^{2})=R$.\\\\

\textbf{Data Availability:} This manuscript has no associated data.

\end{document}